Wall slip in primitive chain network simulations of shear startup of entangled polymers and its effect on the shear stress undershoot

[1*]Yuichi Masubuchi, [2,3]Dimitris Vlassopoulos,

[4]Giovanni Ianniruberto, and [4]Giuseppe Marrucci

[1] *Department of Materials Physics, Nagoya University, Nagoya 4648603, Japan*

[2] *Institute of Electronic Structure and Laser, FORTH, 71110, Heraklion, Crete, Greece*

[3]*Department of Materials Science, Technology, University of Crete, 71003, Heraklion, Crete, Greece*

[4]*Dipartimento di Ingegneria Chimica, dei Materiali e della Produzione Industriale, Università degli Studi di Napoli "Federico II", Piazzale Tecchio 80-80125 Napoli, Italy*

*To whom correspondence should be addressed: mas@mp.pse.nagoya-u.ac.jp

ABSTRACT

In some recent experiments on entangled polymers of stress growth in startup of fast shear flows an undershoot in the shear stress is observed following the overshoot, i.e., before approaching the steady state. Whereas tumbling of the entangled chain was proposed to be at its origin, here we investigate another possible cause for the stress undershoot, i.e., slippage at the interface between polymer and solid wall. To this end, we extend the primitive chain network model to include slip at the interface between entangled polymeric liquids and solid walls with grafted polymers. We determine the slip velocity at the wall, and the shear rate in the bulk, by imposing that the shear stress in the bulk polymers is equal to that resulting from the polymers grafted at the wall. After confirming that the predicted results for the steady state are reasonable, we examine the transient behavior. The simulations confirm that slippage weakens the magnitude of the stress overshoot, as reported earlier. The undershoot is also weakened, or even disappears, because of a reduced coherence in molecular tumbling. In other words, the disentanglement between grafted and bulk chains, occurring throughout the stress overshoot region, does not contribute to the stress undershoot.

INTRODUCTION

As it has been known for a long time, entangled polymers exhibit shear thinning, i.e., at sufficiently large shear stresses the steady-state viscosity decreases with increasing shear rate[1–3]. It is also well known that, during startup of fast shear flows, the transient shear stress shows an overshoot before reaching the steady state[4–6]. Birefringence measurements[5] and molecular simulations[7–9] revealed that the primary molecular mechanism of the overshoot is the flow-induced orientation of entangled subchains between consecutive entanglements[10]. Chain stretch further contributes to the overshoot when the shear rate exceeds the reciprocal Rouse time. Following such well-known overshoot, an undershoot has been observed in some cases[11–13]. Because the undershoot only appears at high shear rates, its origin might be found in either slip at the wall, edge fracture, or other instabilities, and not necessarily in some molecular mechanism. Possibly because of such uncertainties, the undershoot has not been frequently discussed in the literature. Recently, however, rheometry developments allow for more reliable measurements at high shear rates[11,12], and a focus on the undershoot now appears appropriate.

Costanzo et al.[11] attributed the undershoot to the tumbling motion of the entangled polymer chain under shear, such unexpected tumbling having been revealed by recent molecular simulations[14–16]. The effect of the tumbling motion is mathematically described by a damped sinusoidal function. Indeed, a fast shear startup initially triggers a coherent tumbling of all molecules, but subsequently such coherence decays exponentially. The model reasonably reproduces the stress undershoot for some polystyrene melts and solutions[11]. Masubuchi et al.[17] run multi-chain sliplink simulations, confirming the initial coherence of tumbling under startup of fast shear. Stephanou et al.[18] proposed a different molecular theory (the so-called tumbling-snake model), in which the undershoot is also attributed to molecular tumbling.

On the other hand, the undershoot could be linked to other phenomena, such as wall slip[19,20], or even periodic slippage (stick-slip[21–24]), although the data do not indicate the existence of damped periodic response. To examine the effect of transient slippage on the stress growth under shear, Pearson and Petrie[25] proposed a retarded slip boundary condition, in which they introduced the slip relaxation time, reflecting the dependence of slip on the stress history at the wall. Kazatchkov and Hatzikiriakos[26] extended this idea to a multi-mode memory function. Using a K-BKZ constitutive equation, they reproduced the transient viscosity under slippage conditions quantitatively and determined the relaxation times for the slip dynamics of a commercial linear low-density polyethylene melt from LAOS experiments. Ebrahimi et al.[27] used another K-BKZ constitutive model to reproduce the data of a high-density polyethylene melt. However, a single slippage relaxation time was used by them, and chosen close to the peak time of the overshoot. In parallel to the constitutive modeling, a stochastic approach was also used by Hatzikiriakos and Kalogerakis[28], who developed a network model to calculate the configuration distribution function resulting from creation and destruction of transient bonds between the bulk and the wall. They demonstrated that the slippage reduces the magnitude of the stress overshoot, as well as that of the steady-state viscosity.

One might envisage that a transient slippage induces the stress undershoot during the stress decrease from the overshoot, i.e., in the approach to the steady state. It should be mentioned, however, that no undershoot was detected in the studies mentioned in the previous paragraph. In this work, we systematically investigate the possible role of slip at the wall in inducing a stress undershoot in transient shear by running multi-chain slip-link simulations that include slip at the wall. Indeed, simulations allow for more detailed observables than experiments do. To that purpose, we modified the well-established primitive-chain-network (PCN) simulation code by introducing parallel solid plates confining a slab of bulk polymers, and by grafting other polymers at the wall. In the simulation, one of the plates starts moving with a constant velocity, thus generating a shear flow. The plate motion propagates to the bulk liquid, but a slip velocity is also introduced. The latter is determined by imposing that the shear stress is the same at the wall and in the bulk. Similarly to the earlier studies[25–28] mentioned above for the effect of slippage on the viscosity, we here examine the stress response during the entire transient shearing, from startup up to the steady state. Simulations are run for different values of shear rate, bulk chain molar mass, and grafted chain density and molar mass. Effects of the simulation box thickness are also examined.

MODEL AND SIMULATIONS

The simulation code used here is the same as that employed in earlier studies for bulk systems[7,17,29–32], augmented for the presence of parallel walls, at which slip can take place. In the simulation, the bulk entangled polymeric system is replaced (as usual) by a network consisting of strands, nodes, and dangling ends. Each polymer chain corresponds to a path connecting dangling ends through strands and nodes. At each node, two polymer chains are connected by a sliplink through which the two chains are allowed to slide along their backbone, the sliplink somehow restricting the lateral motion. Sliplinks are removed when a chain end goes through them by either reptation or fluctuation. Conversely, a new sliplink is created on the dangling end (by "hooking" a surrounding chain) when the dangling end itself is long enough. The state variables are the sliplink position $\{\mathbf{R}\}$, the number $\{n\}$ of Kuhn segments in each strand and in the dangling ends, and the number $\{Z\}$ of strands (including dangling ends) in each chain. The sliplink positions $\{\mathbf{R}\}$ obey a Langevin-type equation of motion, i.e., a force balance involving the drag force, the strand tensions, an osmotic force and a random one. A force balance also controls the rate of change of $\{n\}$, describing chain sliding through sliplinks. Units of length, energy, and time are the average strand length $a$ at equilibrium, the thermal energy $kT$, and the diffusion time of the node $\tau = \zeta a^2/6kT$, with $\zeta$ the friction coefficient of the node (resulting from the four half-strands emanating from it). In the previous studies for bulk polymers, periodic boundary conditions were used for a cubic simulation box, and flow was applied through a small affine deformation at each integration timestep. Simulations with this model reproduced linear and non-linear rheology of entangled polymers semi-quantitatively[30,31,40–44,32–39].

In this study, we introduce wall boundary conditions as schematically represented in Fig. 1. To mimic confinement between parallel solid walls, we set reflective boundary conditions in the shear-gradient direction, while periodic boundary conditions are maintained in the shear and vorticity directions. Following earlier studies[45–48], we placed tethered polymers at the wall (i.e., with one grafted chain end) with a grafting number density $\rho_g$. To simulate a shear flow, we adopted the procedure detailed below.

We move the "upper" wall with a constant velocity $v_w$. Hence, the nominal shear rate $\dot{\gamma}_n$ is defined as $v_w/d$, where $d$ is the distance between the walls. In a real experiment, the motion propagates to the bulk via the momentum transfer. However, in Brownian dynamics simulations of the bulk motion, we must introduce (as mentioned above) a small shear deformation at each integration timestep. In the present simulation we maintain the assumption that the shear rate is uniform throughout the bulk (no banding), yet, because of the slip at the wall, the rate $\dot{\gamma}_b$ of such deformation is not *a priori* known, except for the obvious inequality $\dot{\gamma}_b \leq \dot{\gamma}_n$.

Hence, the velocity field imposed in the present simulation is given by (see Fig.1)

$$v_x(y) = v_s + \dot{\gamma}_b y \qquad (1)$$

where the subscript $x$ stands for the shear direction, $y$ is the distance from the "lower" wall, and both the slip velocity $v_s$ and the bulk shear rate $\dot{\gamma}_b$ are *a priori* unknown. We determined these two unknowns at each simulation timestep by imposing the following two conditions. First, the shear stress at the wall $\sigma_w$ (generated by the tethered segments under the slip velocity $v_s$ and the bulk shear rate $\dot{\gamma}_b$) and that in the bulk $\sigma_b$ (only generated under the bulk shear rate $\dot{\gamma}_b$) must be set equal to one another:

$$\sigma_w = \sigma_b \qquad (2)$$

The second condition is that the bulk velocity at the upper wall (obeying eq 1) must be smaller than $v_w$ by the same slip velocity $v_s$, implying that

$$v_s = (v_w - \dot{\gamma}_b d)/2 \qquad (3)$$

where $v_s$ and $\dot{\gamma}_b$ are linked again. Note here that we have assumed that $v_s$ is the same at the upper and lower walls (see again Fig. 1).

Note finally that a solution of the above problem (eqs. 2 and 3), leading to the determination of $\dot{\gamma}_b$ and $v_s$, does not always exist. Such a situation appears for sufficiently large values of either the grafted chain density $\rho_g$, or of the grafted chain length $Z_g$, with increasing values of the nominal shear rate $\dot{\gamma}_n$. Indeed, by increasing $\dot{\gamma}_n$ a situation is reached where, even if $v_s$ is set to the value that minimizes the drag exerted on the grafted chains, and hence minimizes the wall shear stress $\sigma_w$, the latter would still be larger than that in the bulk $\sigma_b$, which is clearly impossible. In the real world, under those conditions, slip at the wall would be replaced by shear banding, with a layer of low shear rate close to the wall, and one of high shear rate in the bulk. Dealing with simulations accounting for shear banding is beyond the scope of this work. Work on shear banding is in progress by one of the authors of the present paper by using the Dissipative Particle Dynamics (DPD) technique[49,50].

We performed simulations for several grafting densities $\rho_g$ and molecular weights $Z_g$ of the grafted chains. Unless stated differently, the molecular weight $Z_b$ of the bulk polymers was fixed at 40, and the simulation box size $d$ at 20. Note that molecular masses are indicated in terms of number of entangled segments at equilibrium. For simplicity, both grafted and bulk chains are taken to be monodisperse. The viscoelastic longest relaxation time and the Rouse time for the bulk chains are obtained from the "classical" simulations (without walls), and they come out as $\tau_d = 5.5 \times 10^3$ and $\tau_R = 81$, respectively. For each parameter choice, eight independent simulations were run, starting from different initial configurations. After a sufficiently long equilibration, motion of the upper wall was started. The nominal shear rate $\dot{\gamma}_n$ was varied from 0.001 to 0.3. This range corresponds to $5.5 \leq Wi \leq 1.6 \times 10^3$, and $8.1 \times 10^{-2} \leq Wi_R \leq 24$, which is very similar to the experimental range of Weissenberg numbers in the earlier study[11]. Flow-induced friction reduction[31,51] was ignored in this study.

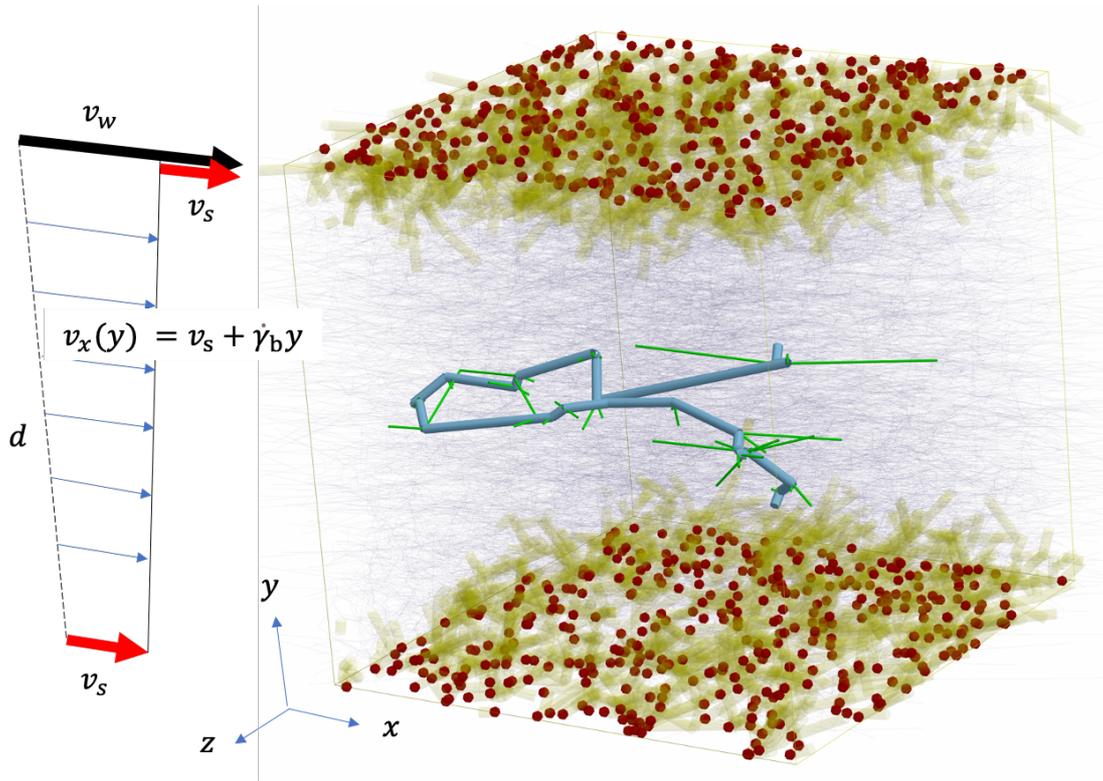

**Figure 1** Schematic representation of the simulation box, with bulk chains in blue and grafted ones in yellow. For better clarity, one of the bulk chains is highlighted, together with its entangled partner segments (in green). Red dots indicate grafted chain ends. For given values of all parameters, the bulk shear rate $\dot\gamma_b$ and the slip velocity $v_s$ are determined as detailed in the text. In this figure, the molecular weights of the bulk and grafted chains are $Z_b = 40$ and $Z_g = 5$, respectively, the grafting density is $\rho_g = 2$, and the box dimension is $d = 20$.

RESULTS AND DISCUSSION

Before showing steady shear results in the nonlinear range, it is convenient to report equilibrium results on grafted chains. Figure 2 shows the longest relaxation time $\tau_g$ of the grafted chains (entangled with the bulk chains having $Z_b = 40$) for several values of the grafted-chain density $\rho_g$ and molar mass $Z_g$. Such relaxation time was determined from the grafted-chain stress autocorrelation function. In Fig 2 (a), $\tau_g$ is shown to increase with increasing $Z_g$, as expected from the arm-retraction relaxation mechanism. In Fig 2 (b), $\tau_g$ is shown to decrease with increasing $\rho_g$, seemingly because entanglements between grafted chains increase, and correspondingly those between bulk and grafted chains decrease (results not shown).

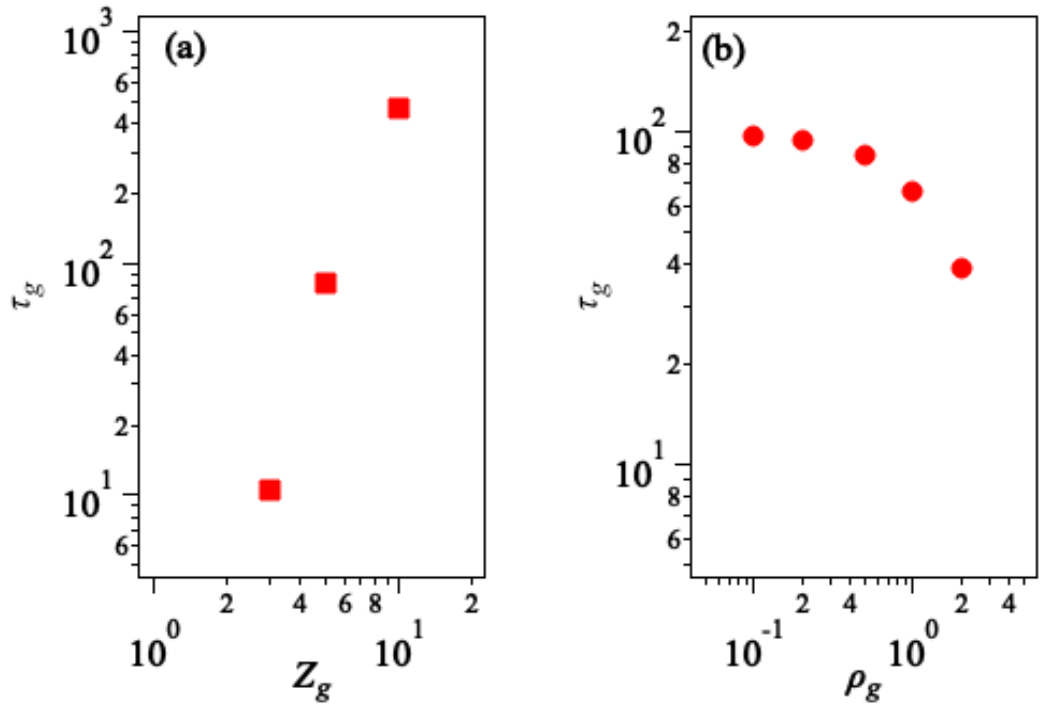

**Figure 2** Relaxation time $\tau_g$ of the grafted chains for the case $Z_b = 40$, as a function of $Z_g$ (left) and $\rho_g$ (right).

We now move on to show in Fig 3 the steady-state shear stress $\sigma = \sigma_w = \sigma_b$ plotted against the nominal shear rate $\dot{\gamma}_n$, for several values of the grafted-chain molar mass $Z_g$ (left) and density $\rho_g$ (right), all for a bulk-chain molar mass $Z_b = 40$. For comparison, the flow curve obtained from the bulk simulations without walls is also shown (black filled circles).

The arrows in Fig 3 (a) indicate the values of $1/\tau_g$ taken from panel (a) of Fig 2. Hence, Fig 3 (a) shows that, at the fixed value $\rho_g = 0.5$, and for nominal shear rates $\dot{\gamma}_n$ larger than $1/\tau_g$, the shear stress (for all reported $Z_g$ values) essentially coincides with that of the simulations without walls. Conversely, when $\dot{\gamma}_n < 1/\tau_g$ the stress runs significantly below, revealing strong wall slip effects. In this low-shear-rate range, grafted chains have all the time to disentangle from the bulk ones, hence making bulk molecules free to slip away from the wall. We can call such behavior thermally-induced slip.

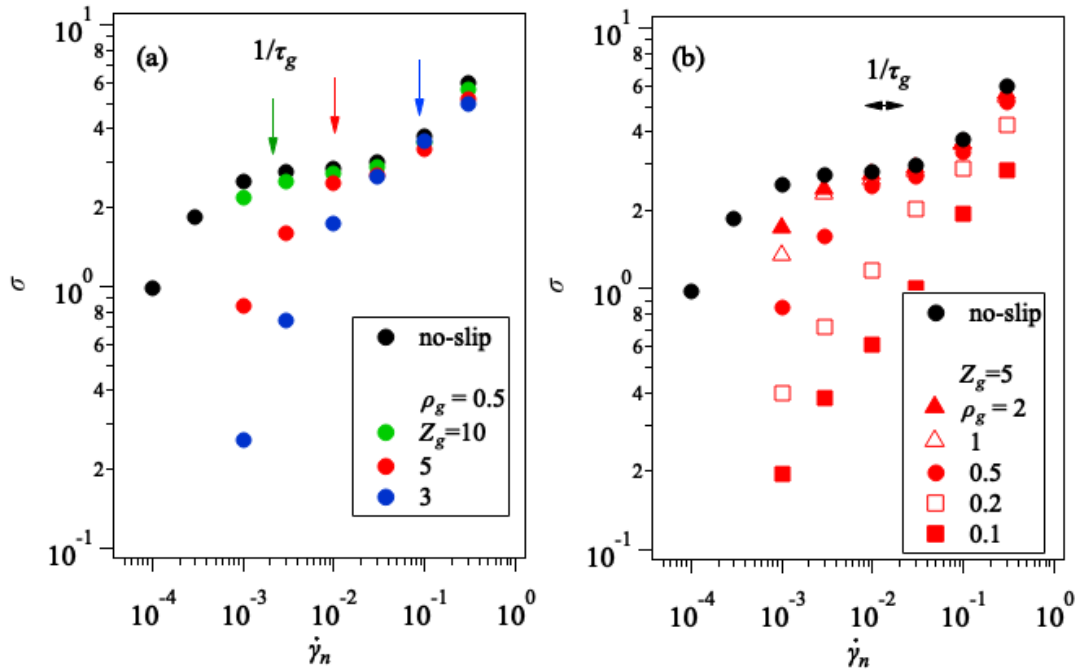

**Figure 3** Shear stress vs. nominal shear rate for $Z_b = 40$. The left panel (a) is for $\rho_g = 0.5$ and various $Z_g$, while the right panel (b) is for $Z_g = 5$ and various $\rho_g$. Results for the no-slip case are also shown as black circles. Arrows indicate the relaxation rate $1/\tau_g$ of the grafted chains.

The just described situation only applies at large enough grafted-chain densities.

Indeed, Fig 3 (b) shows that, at the fixed value $Z_g = 5$, a large slip occurs also for $\dot{\gamma}_n > 1/\tau_g$ (the horizontal arrow showing the $1/\tau_g$ range $0.01 \div 0.02$) provided the density $\rho_g$ is sufficiently small. For $\dot{\gamma}_n > 1/\tau_g$, slip is no longer thermally induced, but rather it is due to convection (flow-induced slip). Such flow-induced slip is clearly visible at low grafted-chain densities.

We conclude the discussion of Fig 3 by noting that, if larger values of $Z_g$ are considered (like $Z_g = 20$, or even $Z_g = 40$), the thermally-induced slippage is suppressed in a sensible range of shear rates because $\tau_g$ becomes too large (results not shown). At the same time, however, also the flow-induced slippage is suppressed because the condition $\sigma_w = \sigma_b$ cannot be fulfilled, as mentioned in the previous section. For such case, shear banding (instead of slip at the wall) prevails.

Figure 4 shows the slip velocity $v_s$ plotted against the nominal shear rate $\dot{\gamma}_n$. The straight line in both panels represents the maximum value of $v_s$ given by $v_{s,max} = \frac{1}{2}v_w = \frac{1}{2}\dot{\gamma}_n d = 10\dot{\gamma}_n$. Before discussing Fig 4, it is appropriate to recall the earlier experimental study of Durliat et al.[52] showing a slip transition at a critical shear rate, at which the value of $v_s$ abruptly increased by more than one decade. Both before and after such transition, the slip velocity grows linearly with the shear rate [52]. In our simulations, for sufficiently high values of $\rho_g$ and $Z_g$, the slip velocity $v_s$ steeply increases at values of $\dot{\gamma}_n$ somewhat smaller than 0.1 (see green circles in the left panel, and filled and unfilled triangles in the right one). Although the observed transition is not as sharp as experimentally reported, probably due to a size effect (our simulation "sample" is very thin), our results appear consistent with the experiments of Durliat et al.[52]. Moreover, after the transition, i.e., to the right of $\dot{\gamma}_n \approx 0.1$, our simulation results appear to approach a unit slope, consistently with Durliat et al.[52].

When the thermally-induced slip dominates, i.e., at $\dot{\gamma}_n < 1/\tau_g$, the slip velocity $v_s$ runs very close to its maximum value, $v_{s,max}$. This is apparent in the leftmost results in both panels of Fig 4. Conversely, for $\dot{\gamma}_n > 1/\tau_g$ we observe the transition to flow-induced slip, where the slip velocity $v_s$ is significantly lower than $v_{s,max}$. Such transition is particularly evident in the red and blue circles in Fig 4, whereas it falls too low to be observed for large values of either $\rho_g$ or $Z_g$. No transition at all is observed for low values of $\rho_g$ where, in spite of switching from the thermally-induced to the flow-induced slippage, the slip velocity remains essentially coincident with $v_{s,max}$ (see red squares in panel b). Indeed, at low densities, the grafted-chain stress is obviously very small, and hence the equal value of stress in the bulk polymer implies a very small value of the bulk shear rate $\dot{\gamma}_b$. For $\dot{\gamma}_b \approx 0$, from Eq 3 we then get: $v_s \approx \frac{1}{2} v_w = v_{s,max}$.

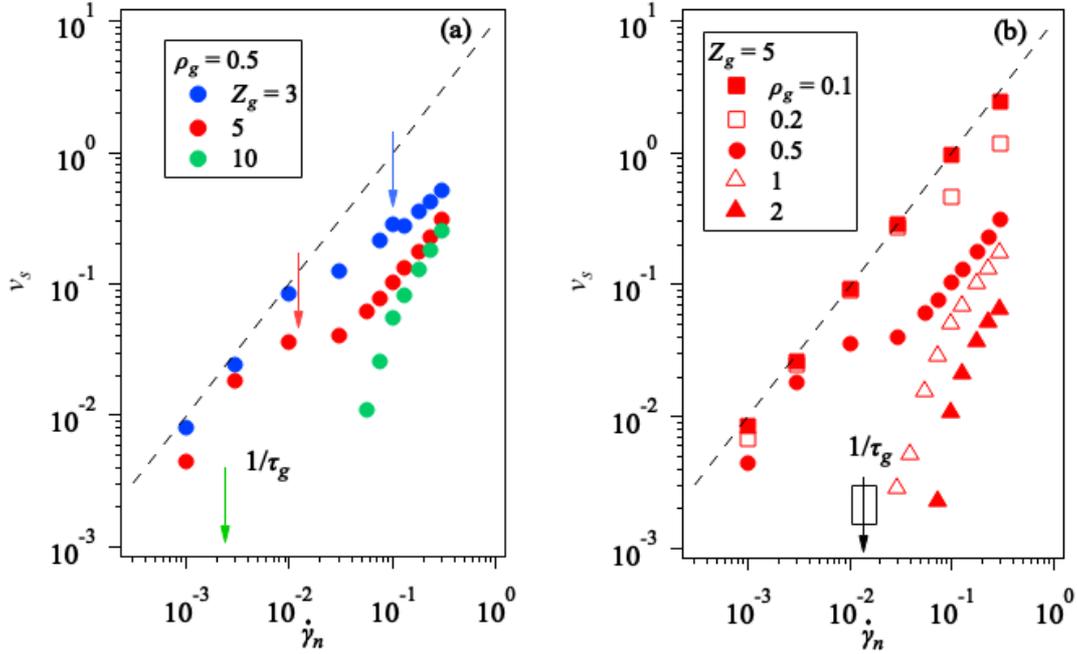

**Figure 4** Slip velocity $v_s$ against nominal shear rate for $Z_b = 40$. The left panel (a) is for $\rho_g = 0.5$, and several lengths $Z_g$ of the grafted chains, while the right one (b) is for $Z_g = 5$, and several $\rho_g$ values. The straight lines give the maximum possible value of $v_s$.

Figure 5 shows the slip velocity $v_s$ as a function of the shear stress $\sigma$ for several $Z_g$ (left panel) and $\rho_g$ (middle panel) values. The slip velocity significantly increases with increasing the shear stress, beyond a critical value $\sigma_c$, which depends on $Z_g$ and $\rho_g$. As noted by Hatzikiriakos[53], the $v_s$ vs. $\sigma$ curves superpose on each other if $v_s$ is plotted against $\sigma/\sigma_c$ (right panel in Fig. 5). Note that we determined $\sigma_c$ from the behavior at high $\sigma$ values, i.e., by ignoring the data in the region of thermally-induced slippage. Specifically, we estimated $\sigma_c$ in such a way that the normalized data superimpose for $\dot{\gamma}_n$ larger than $1/\tau_g$. For instance, for the case $Z_g = 3$ (blue circles in the mid panel of Fig. 5), we determined $\sigma_c$ by ignoring results for $\sigma < 3$, where the slope is clearly different from that in the high $\sigma$ region insofar as results are still sensitive to thermally-induced slippage.

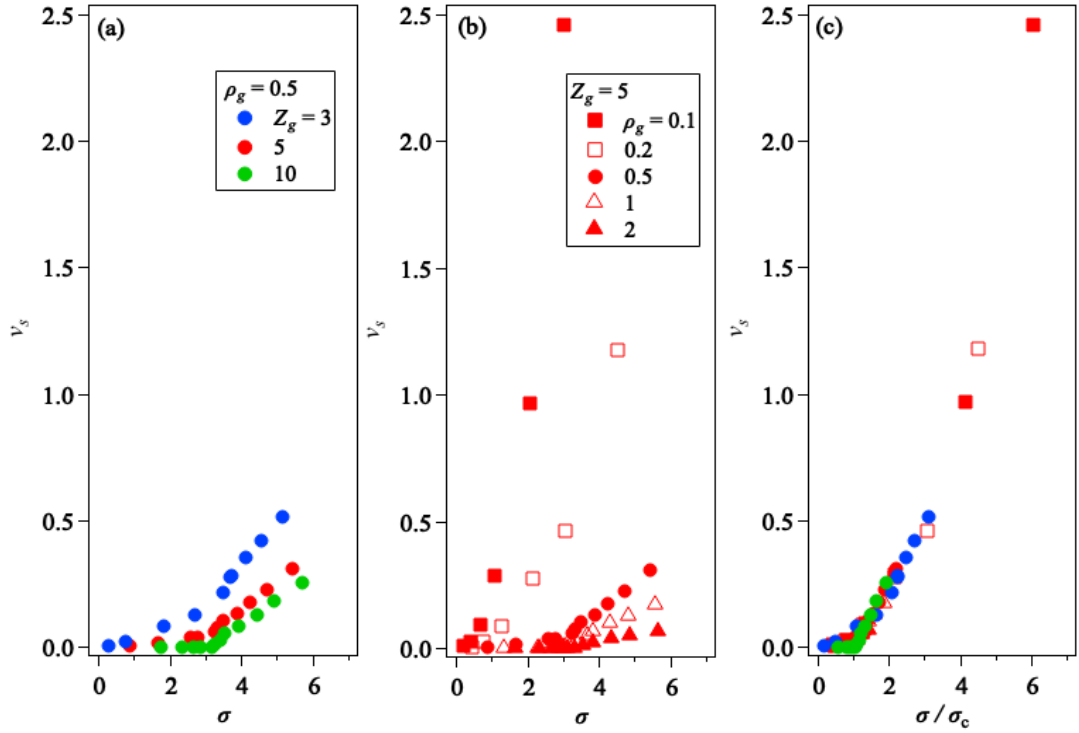

**Figure 5** Slip velocity $v_s$ against shear stress $\sigma$ for several $Z_g\rho_g$ (left) and $Z_g\rho_g$ (middle) values. The right panel shows $v_s$ plotted against the normalized shear stress $\sigma/\sigma_C$.

Figure 6 reports the viscosity growth curves for several $\rho_g$ and $Z_g$ values. Here, the "apparent" viscosity was calculated as $\eta_a = \sigma/\dot{\gamma}_n$. For comparison, the results from the no-slip simulations are also shown in Fig. 6 as broken curves. As previously shown in Fig. 2, Fig. 6 confirms that at low $\dot{\gamma}_n$ the stress (hence $\eta_a$) is lower than that in the no-slip case, due to the thermally-induced slippage. At higher $\dot{\gamma}_n$, larger than the disengagement rate of the grafted chains shown in Fig 3, the simulation results with and without slippage are close to one another (provided $\rho_g$ is large enough), as shown by way of example in Fig. 6(c). In these high-$\dot{\gamma}_n$ flows, the viscosity growth curve depends on $\rho_g$ and $Z_g$.

Panels from (a) to (c) in Fig. 6 show results at the fixed $Z_g = 5$ value, for three different values of $\rho_g$. As previously mentioned, the steady-state viscosity increases with increasing $\rho_g$ due to suppression of slippage. Concerning the transient behavior, the viscosity overshoot is reduced at low $\rho_g$ values. In fact, for $\rho_g = 0.1$ (panel a) there is no overshoot at all, except perhaps at the highest shear rate. For $\rho_g = 0.5$ (panel b), an overshoot is observed for $\dot{\gamma}_n \geq 0.03$, but the peak value is reduced, and its position is retarded, compared to the no-slip case. For $\rho_g = 2$, and at high rates, there is no difference between slip and no-slip simulations.

The effect of $Z_g$ can be seen in panels (b), (d), and (e) of Fig. 6, where $\rho_g$ is fixed at 0.5, while $Z_g$ takes the values 5, 3 and 10, respectively. These plots reveal that the slippage in slow flows is suppressed for large $Z_g$ because grafted chains remain entangled. Indeed, for $\dot{\gamma}_n < 0.01$, the steady-state viscosity decreases with decreasing $Z_g$, as already seen in Fig. 2. Conversely, in flows faster than the disengagement rate of grafted chains, the steady-state viscosity is almost independent of $Z_g$, and close to the no-slip value, because slippage does not occur. On the other hand, the transient behavior is somewhat sensitive to $Z_g$. By way of example, let's compare the orange curves ($\dot{\gamma}_n = 0.01$) in panels (b), (d), and (e), where the overshoot appears very different. Namely, for $Z_g = 10$ (panel e) the overshoot is clearly visible, though the peak value is reduced, and the peak position is retarded, with respect to the no-slip case; for $Z_g = 5$ (panel b) the overshoot is sort of truncated; finally, for $Z_g = 3$ (panel d) the overshoot has completely disappeared.

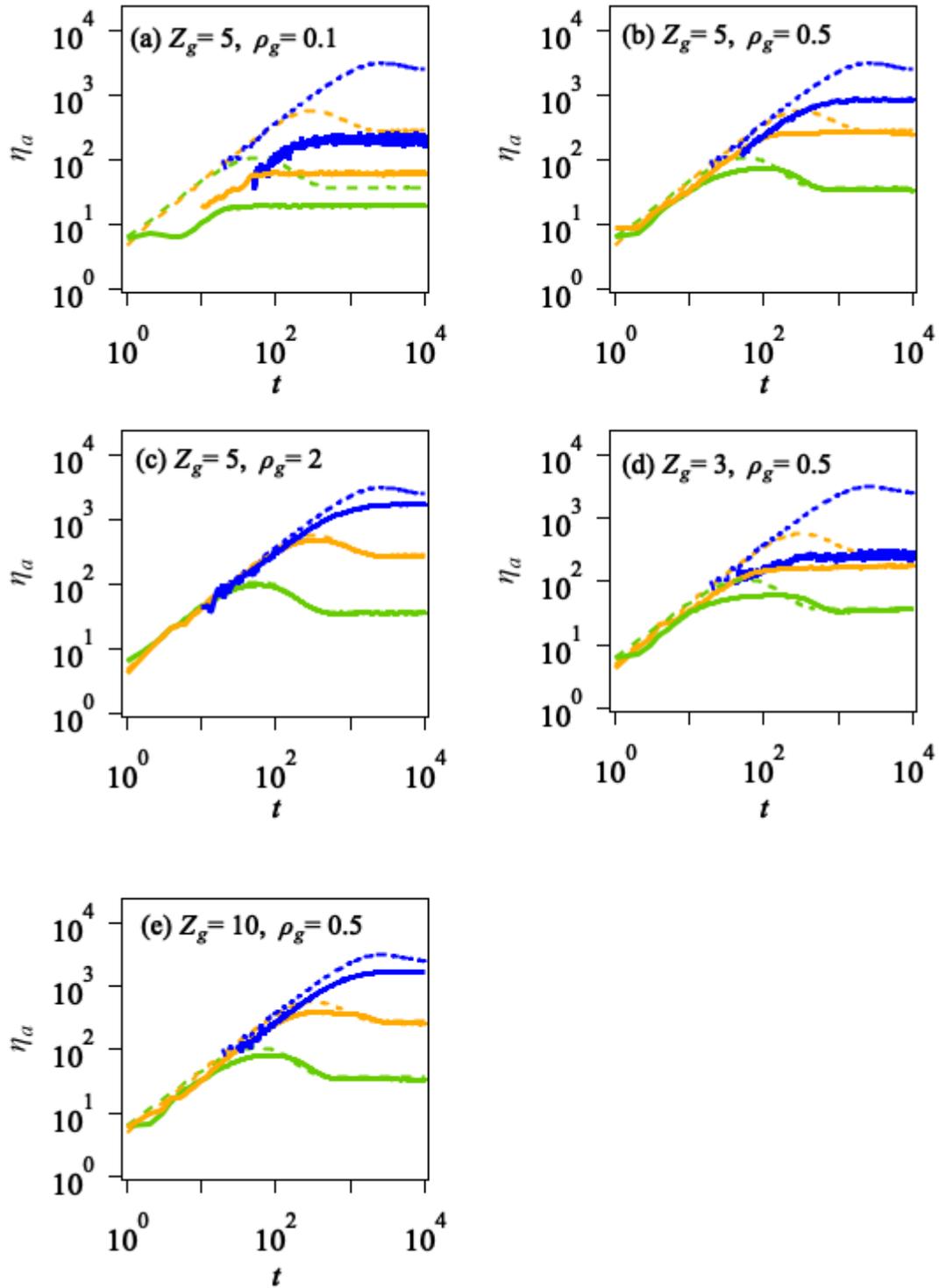

**Figure 6** Viscosity growth curves for several $\rho_g$ and $Z_g$ values. Nominal shear rates are 0.001 (blue), 0.01 (orange), and 0.1 (green). Broken curves indicate results from the no-slip simulations.

Figure 7 shows further details of the transient behavior for $\dot{\gamma}_n = 0.03$ and $\rho_g = 0.5$, and for several $Z_g$ values. As mentioned above, the stress overshoot is reduced for short grafted chains (see Fig. 7a). Transient values of the slip velocity $v_s$ are shown in Fig. 7(b). Here, $v_s$ first grows with time, and then, after a peak, it decreases towards the steady value. The time development of $v_s$ is similar to that of $\sigma$, as predicted by the stochastic model of Hatzikiriakos and Kalogerakis[28]. The decreasing branch of the $v_s$ and $\sigma$ curves moves to longer times as $Z_g$ decreases. This trend might seem counterintuitive in view of the fact that the relaxation time of the grafted chains decreases with decreasing $Z_g$. However, the delayed development of the stress with decreasing $Z_g$ is in fact due to the correspondingly smaller shear rate in the bulk, $\dot{\gamma}_b$. It is well known that in the shear startup of entangled polymers the orientation-induced stress peak is located at a shear deformation of ca. 2.3[10], and at higher deformation values when also chain stretch comes into play ($Wi_R > 1$). The corresponding peak times are then obtained from the ratio of the peak shear deformation to the shear rate. The above is true, however, only when slip is absent. Conversely, when slip is present, the time at which such an effective deformation is reached, i.e., the peak time, becomes longer the smaller is $\dot{\gamma}_b$.

The number of entanglements formed at the interface between bulk and grafted chains, called $Z_{int}$, is shown in Fig. 7(c). $Z_{int}$ decreases with time from the equilibrium value due to the flow-induced disentanglement between bulk and grafted chains. At long times, the rates of creation and destruction of entanglements at the interface balance one another, and $Z_{int}$ reaches a steady value. Before reaching such steady state, however, $Z_{int}$ shows a clear undershoot (see panel c). As expected, the position of the undershoot (i.e., the maximum disentanglement at the interface) coincides with that of the overshoot of $v_s$ and $\sigma$.

In panels (d)-(f) of Fig. 7, the stress is decomposed (according to the decoupling approximation) into averages $S$ of segment orientation, $\lambda^2$ of square segment stretch ratio, and $Z/Z_0$ of normalized entanglement density. Both with and without slippage, the stress overshoot is essentially determined by the segment orientation $S$ (panel d), whereas the segment stretch $\lambda^2$ comes into play after the overshoot (panel e). Indeed, at the shear rate of 0.03 in Fig. 7 the stretch is also important because the Rouse-based Weissenberg number is larger than unity ($Wi_R = 2.4$). Flow-induced reduction of the entanglement density $Z$ is also observed (panel f). For all these quantities, the curves are essentially shifted to longer times with decreasing $Z_g$, due to the reduced shear rate in the bulk. Changes in the steady values are also consistent with change in the bulk shear rate.

Finally, panel (g) of Fig. 7 shows the effect of molecular tumbling through the average of $\cos^2\theta$, where $\theta$ is the tilt angle of the end-to-end vector of bulk chains to the shear direction. As reported previously[17], in the no-slip case $\cos^2\theta$ decreases over time, and exhibits an undershoot before reaching the steady state. This undershoot reflects a coherent molecular tumbling at startup of the shear flow, and it probably causes an undershoot of $\sigma$ (as assumed by Costanzo et al.[11]), though very weak in panel (a) of Fig.7. The undershoot of $\cos^2\theta$ is suppressed by the slippage, due to reduction of the bulk shear rate $\dot{\gamma}_b$. It so appears that slip mitigates molecular tumbling, presumably by inducing a stronger loss of coherence. Note also that in the present simulations with confining walls, the value of $\cos^2\theta$ at short times is smaller than that in simulations without walls because of the oriented chains near the surface.

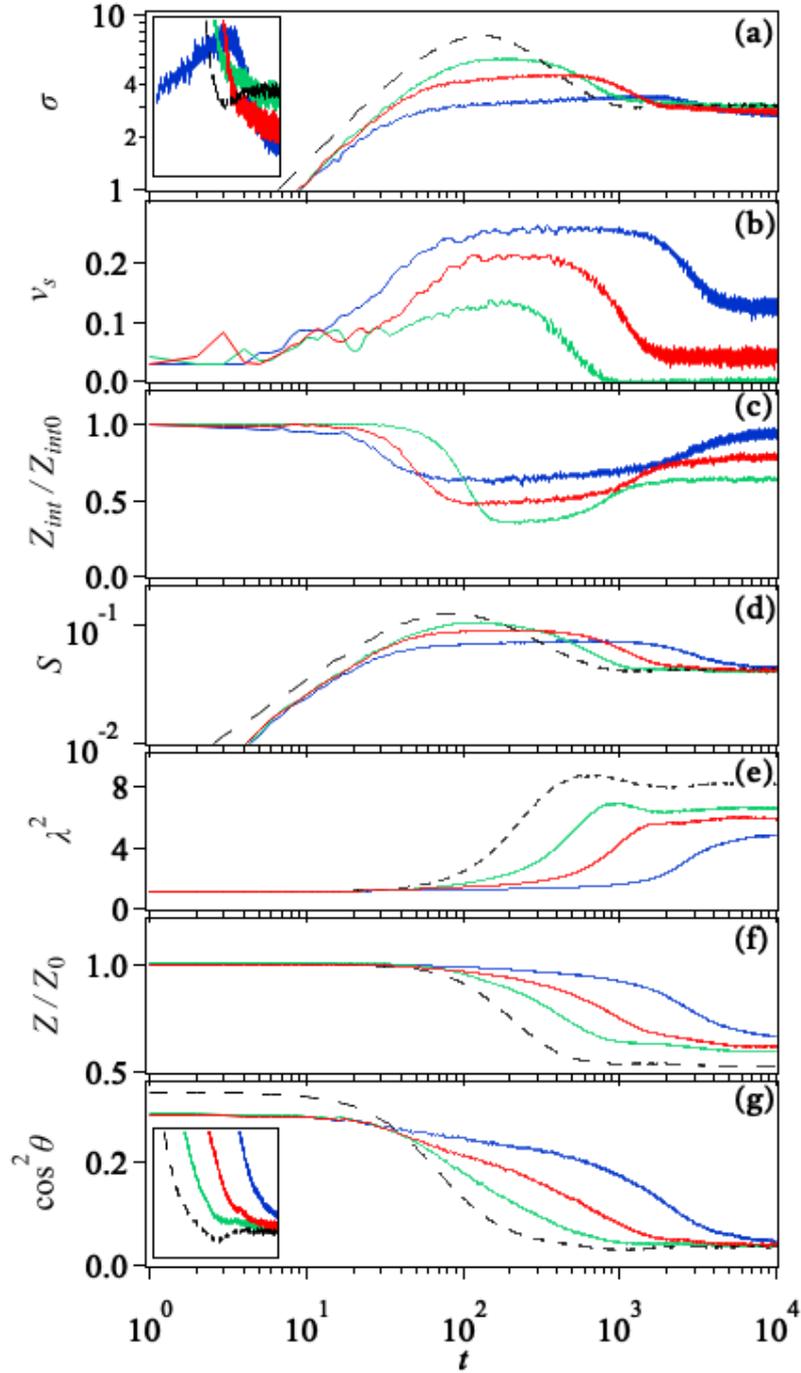

**Figure 7** Shear startup behavior at $\dot{\gamma}_n = 0.03$ ($Wi = 1.6 \times 10^2$ and $Wi_R = 2.4$) and $\rho_g = 0.5$ for $Z_g = 3, 5, 10$ (blue, red, green, respectively). (a) Shear stress, (b) slip velocity, (c) interfacial entanglement density, (d) segment orientation, (e) segment stretch, (f) bulk entanglement density, (g) tilt angle. Dashed curves are from simulations without slip. Insets in panels (a) and (g) are magnified plots in the range $10^2 < t < 10^4$ to show undershoots in the no-slip case (black curves).

Figure 8 shows a comparison among different bulk molecular weights $Z_b$ at a fixed shear rate of $\dot{\gamma}_n = 0.03$, and for $\rho_g = 0.5$ and $Z_g = 5$. At this rate, the steady state shear stress appears insensitive to $Z_b$, and also the slip velocity $v_s$ and interfacial entanglement density $Z_{int}$ do not vary much. Conversely, the transient responses are significantly affected by $Z_b$. Namely, the maximum value of the slip velocity $v_s$ increases with increasing $Z_b$, and the growth of $v_s$ reduces the stress overshoot. The magnitude of the undershoot in the interfacial entanglement density $Z_{int}$ increases with increasing $Z_b$, and the recovering of $Z_{int}$ is retarded at large $Z_b$. This behavior of $v_s$ and $Z_{int}$ implies that the flow-induced slippage is enhanced for longer bulk chains. Notice, however, that no undershoot is observed in the $\sigma$ curves.

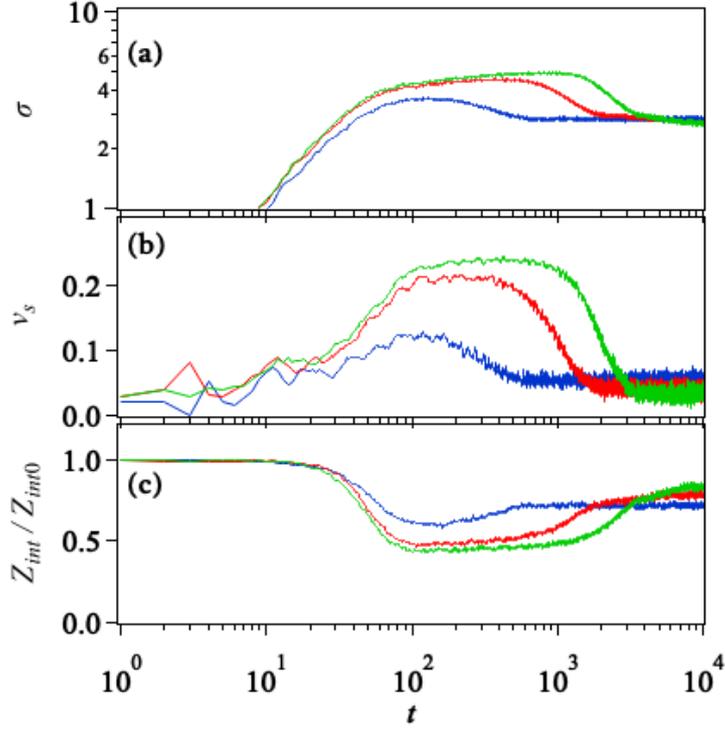

**Figure 8** Shear startup behavior of (a) shear stress, (b) slip velocity, and (c) interfacial entanglement density, for $Z_b = 20$ (blue), 40 (red) and 60 (green) with $\rho_g = 0.5$, $Z_g = 5$, and $\dot{\gamma}_n = 0.03$.

Finally, Fig. 9 shows the box-size effect, i.e., the effect of changing the thickness $d$ of the sheared layer on the same quantities of Fig. 8. However, the slip velocity is here plotted in normalized form, i.e., as the sum of the upper and lower slip velocities divided by the wall velocity. Such ratio, $r = 2\, v_s/v_w$, is here appropriate because, for an equal value of $\dot{\gamma}_n$, changing $d$ proportionally increases $v_w$, and hence (if not necessarily in the same proportion) also $v_s$. It is also worth noting that, in view of eq 3, the above velocity ratio can also be written as $r = 1 - \dot{\gamma}_b/\dot{\gamma}_n$. The normalized slip velocity $r$ varies in the range $0 \div 1$, the lower limit implying a zero slip velocity, while the upper one a zero bulk shear rate (total slip).

Figure 9 shows that box-size effects are minor, but not totally negligible. This is because the slip velocity (as opposed to the normalized one) also plays some role. By increasing the layer thickness $d$, the slip velocity increases, and hence grafted chains are dragged more effectively by the bulk flow; correspondingly, the shear stress somewhat increases (the green curve runs slightly higher in panel a). For the same reason, i.e., because of the increased slip velocity, the flow-induced disentanglement of the grafted chains from the bulk ones increases (the green curve runs slightly lower in panel c). Concerning the stress undershoot, there is no sign of it in all cases here examined.

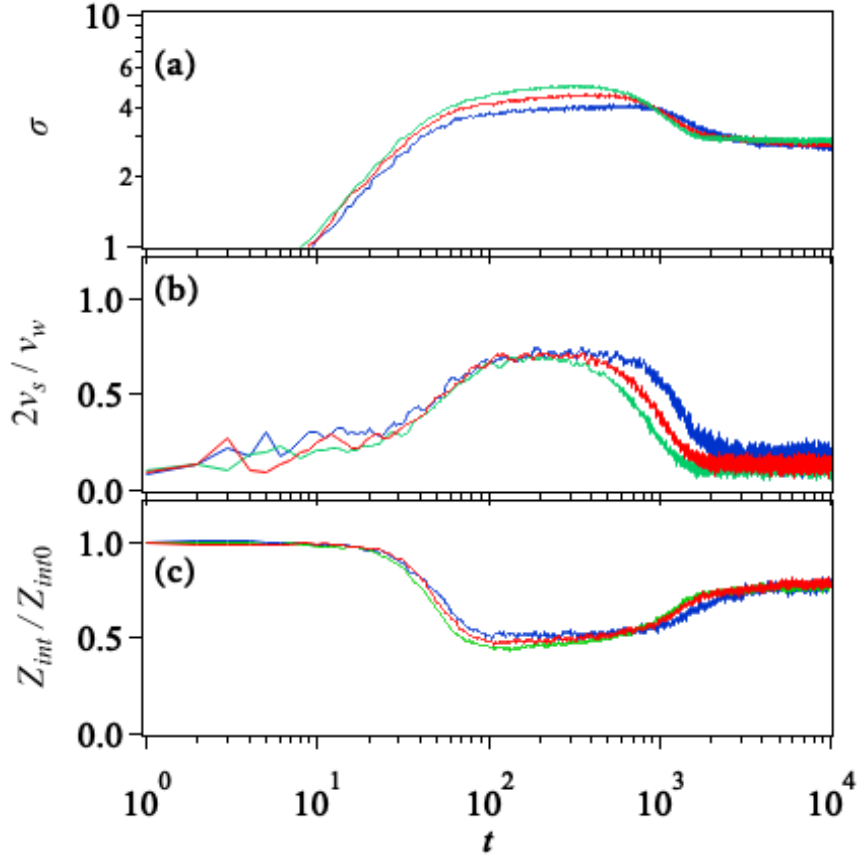

**Figure 9** Effect of changing the box size from $d = 15$ (blue curves), to 20 (red), and 25 (green) on shear startup at the fixed nominal shear rate $\dot{\gamma}_n = 0.03$. The observed quantities are (a) shear stress, (b) normalized slip velocity, and (c) interfacial entanglement density, all of them for $\rho_g = 0.5$, and $Z_g = 5$.

CONCLUDING REMARKS

In this study, we extended the primitive chain network model to deal with slippage between entangled polymeric liquids and solid walls with grafted polymers. We determined the slip velocity and the bulk shear rate by fulfilling the condition that the shear stress developed in the bulk chains and that resulting from the grafted chains must be equal to one another. By varying the molecular weight and the density of the grafted chains, we run several shear startup simulations. Results confirmed that indeed slip takes place if the chains attached at the wall are either short or sparsely grafted. For the transient startup behavior, the stress overshoot was found to be weakened by the slippage. During startup, the effect of slippage becomes more retarded when the grafted chains are longer. These results are consistent with earlier studies, and they essentially validate the model.

On the other hand, the main objective of this study, concerning the possible role of the wall slip in inducing stress undershoot, was fully reached, in the sense that no sign of undershoot is ever found in all simulations performed here. Disentanglement between grafted and bulk chains was found in our simulations in the stress overshoot region. However, such disentanglement did not induce any undershoot. On the contrary, it so appears that the slippage prevents the undershoot because it enhances loss of coherence in the molecular tumbling.

We conclude by mentioning that the Brownian dynamics approach used here is not the only simulation technique able to investigate slip. Work based on the DPD method mentioned earlier[49,50] is now in progress by one of the authors.

ACKNOWLEDGEMENTS

YM was supported in part by Ogasawara foundation, JST-CREST (JPMJCR1992), and NEDO (JPNP16010).